\documentclass[12pt]{article}
\usepackage{subcaption}
\usepackage{setspace}
\usepackage{amsmath, amssymb, amsthm, float, graphicx}
\pdfoutput=1
\usepackage{xcolor}
\usepackage{hyperref}
\usepackage[utf8]{inputenc}
\numberwithin{equation}{section}
\hypersetup{colorlinks=false, linkcolor=blue, citecolor=red}
\def\g{\gamma}
\def\G{\Gamma}

\def\m{\mu}
\def\n{\nu}

\def\e{\epsilon}

\def\b{\beta}
\def\d{\delta}

\def\f{\phi}

\def\D{\Delta}

\def\ta{\tau}

\renewcommand{\b}{\beta}

\renewcommand{\d}{\delta}
\newcommand{\dsl}{\pa \kern-0.5em /}

\newcommand{\half}{\frac{1}{2}}
\newcommand{\pa}{\partial}

\newcommand{\nn}{\nonumber\\}

\begin{document}
\title{\bf More on analytic bootstrap for O(N) models }
\date{}

\author{Parijat Dey\footnote{parijat@cts.iisc.ernet.in},~ Apratim Kaviraj\footnote{apratim@cts.iisc.ernet.in} ~and Kallol Sen\footnote{kallol@cts.iisc.ernet.in} \\ ~~~~\\
\it Centre for High Energy Physics,
\it Indian Institute of Science,\\ \it C.V. Raman Avenue, Bangalore 560012, India. \\}
\maketitle
%\vskip 2cm
%\abstract{In order to achieve a better analytic handle on the modern conformal bootstrap program, we re-examine and extend the pioneering 1974 work of Polyakov's, which was based on consistency between the operator product expansion and unitarity. As in the bootstrap approach, this method does not depend on evaluating Feynman diagrams. We show how this approach can be used to compute the anomalous dimensions of certain operators in the $O(n)$ model at the Wilson-Fisher fixed point in $4-\epsilon$ dimensions up to $O(\epsilon^2)$.}
\abstract{This note is an extension of a recent work on the analytical bootstrapping of $O(N)$ models. An additonal feature of the $O(N)$ model is that the OPE contains trace and antisymmetric operators apart from the symmetric-traceless objects appearing in the OPE of the singlet sector. This in addition to the stress tensor $(T_{\mu\nu})$ and the $\phi_i\phi^i$ scalar, we also have other minimal twist operators as the spin-1 current $J_\mu$ and the symmetric-traceless scalar in the case of $O(N)$. We determine the effect of these additional objects on the anomalous dimensions of the corresponding trace, symmetric-traceless and antisymmetric operators in the large spin sector of the $O(N)$ model, in the limit when the spin is much larger than the twist. As an observation, we also verified that the leading order results for the large spin sector from the $\epsilon-$expansion are an exact match with our $n=0$ case. A plausible holographic setup for the special case when $N=2$ is also mentioned which mimics the calculation in the CFT. }

\tableofcontents
%\tableofcontents

\onehalfspacing

\section{Introduction and summary of results}

The application of the crossing symmetry to finding the critical exponents of the $O(N)$ model in the context of $\e-$expansion, dates back to the seminal work of Polyakov,\cite{polyakov} in 1973 and related works in \cite{ferrara, zamolodchikov}. A recent study in \cite{Sen:2015doa} extended the results of Polyakov for the next to the leading order calculation in the epsilon expansion. Other modern methods have also been been explored \cite{Rychkov:2015naa,Koch,Basu}. On the other hand, a significant amount of work has been done on the conventional bootstrap approach \cite{bootstrap,bootstrap 2}. Of these, notable works regarding the $O(N)$ vector models include \cite{Osborn:1993cr, Petkou:2003zz}. While the main features of these works involved the developments of the conformally invariant OPE for the $O(N)$ models, these works were focussed on the perturbative developments as $1/N$ expansion or the $\e-$expansion for example. But a formal non perturbative development of the conformal bootstrap program for the $O(N)$ models was yet to be developed. With the explicit expressions of the conformal blocks in \cite{Dolan1}, \cite{Dolan2} and the subsequent works, it was possible to analyze the modern and conventional bootstrap numerically to find various bounds on the operator dimensions, central charges and coupling constants ({\it i.e.} the OPE coefficients) as discussed in \cite{Rattazzi:2010yc,Vichi:2011ux,Poland:2011ey} and so on.

Recent numerical studies \cite{Kos:2013tga} have shed more light on the non perturbative regime of the $O(N)$ models where they have showed that it is possible to obtain results of the dimensions of certain operators for finite $N$ case which resembles realistic models {\it e.g} the Ising model. Meanwhile, on the analytical side, the authors of \cite{Fitzpatrick:2012yx} have shown that it is possible to analytically explore a certain regime of the spectrum dominated by the large spin sector of operators. A similar argument regarding the leading twist behaviour of the large spin sector was also forwarded in \cite{Komargodski:2012ek}. The authors showed that with the stress tensor in the spectrum, the bootstrap equation can be satisfied by an infinite tower of large spin operators with twists given by,
\begin{equation}
\Delta=2\Delta_\phi+2n+\ell+\g(n,\ell)\,,
\end{equation}
where $\Delta_\phi$ are the conformal dimensions of the external scalars and $\g(n,\ell)$ are the anomalous dimensions for these operators. For more related works see \cite{Li2,Alday:2015eya,Alday:2015ewa}. While the authors have considered a special case for $n=0$, the subsequent works \cite{Kaviraj:2015cxa}, \cite{Kaviraj:2015xsa} have extended this calculation for the $n\neq0$ case. In these papers, it was shown that it is possible to obtain exact analytical expression for the anomalous dimension in terms of the twist $(n)$ and also that an universal contribution of the anomalous dimension can be extracted in the limit when $\ell\gg n\gg1$ given by a generic form,
\begin{equation}
\g(n,\ell) \propto \frac{n^d}{\ell^{d-2}}\,,
\end{equation}
%where $f(d)$ is the universal coefficient depending on the spacetime dimension $d$ and the central charge of the theory. 

In a more recent work,\cite{Li:2015rfa}, the authors have extended this analytical technique of \cite{Fitzpatrick:2012yx} in the case of the $O(N)$ model for the special case of $n=0$. An additional complication for the O(N) case is that the OPE contains trace, and antisymmetric-traceless objects in addition to the usual symmetric-traceless piece. In general,
\begin{equation}
\phi_i(x)\times\phi_j(y)=\sum_{{\mathcal{O}}_{\Delta,\ell}}C^{\Delta,\ell}_{ijk}{\mathcal{O}}^k_{\Delta,\ell}\,,
\end{equation}
where now, ${\mathcal{O}}^k_{\Delta,\ell}$ include all the operators as trace, symmetric-traceless and antisymmetric-traceless ones. Thus a generic four point function of the fundamentals of the $O(N)$ can be reduced in terms of these tensor structures,
\begin{equation}
\langle\phi_i(x_1)\phi_j(x_2)\phi_k(x_3)\phi_l(x_4)\rangle=\delta_{ij}\delta_{kl}I(u,v)
+(\delta_{ik}\delta_{jl}+\delta_{il}\delta_{jk}-\frac{2}{N}\delta_{ij}\delta_{kl})S(u,v)
+(\delta_{ik}\delta_{jl}-\delta_{il}\delta_{jk})A(u,v)\,,
\end{equation}
where $I(u,v)$, $S(u,v)$ and $A(u,v)$ are the conformal blocks for the exchange of trace, symmetric-traceless and antisymmetric-traceless objects. By imposing a crossing symmetry , the large spin sector of the corresponding trace, symmetric-traceless and antisymmetric-traceless operators in the crossed $(t)$ channel can be written in terms of a linear combination of $I(u,v)$, $S(u,v)$ and $A(u,v)$ appearingin the direct $(s)$ channel. Finally analysing each of these contributions arising from these constraint equations, we can solve for the anomalous dimensions of the operators of the large spin sector in terms of the contributions of these minimal twist operators in the direct channel. 

Before moving on to summarize our findings, we would like to put forward, example set up of a bulk calculation for an $O(2)$ model where the analog of the $O(2)$ scalars on the CFT is a complex charged scalar coupled to an $U(1)$ bulk gauge field and gravity. The effective potential for the scalar in presence of the graviton and the gauge field interaction following \cite{Fitzpatrick:2011hh} is thus given by,
\begin{equation}
V_{eff}[\phi,\phi^\dagger]=V_{qart}[\phi,\phi^\dagger]+\frac{\kappa}{2}A^\m J_\m[\phi,\phi^\dagger]+\frac{\kappa}{4}h^{\m\n}T_{\m\n}[\phi,\phi^\dagger]\,,
\end{equation}
where $V_{quar}[\phi,\phi^\dagger]$ is the scalar interaction. In presence of the graviton and the gauge interactions, the anomalous dimensions for the generalized free fields in bulk $AdS_5$ are given by in terms of shifts in the binding energy, in the semi-Newtonian approximation as an expansion in the inverse distance. We would like to point out that in this analysis we are not assuming a bulk description of the $O(2)$ model itself. Since $O(2)$ is not a large $N$ theory, hence the correct description of the bulk dual is not the classical gravity residing in AdS but the full type II B superstring theory in ten dimensions. Instead we are assuming that we still have a large $N$ theory with a gravity dual and the $O(2)$ model is a perturbation on this large $N$ theory. Also the deformation introduced by the $O(2)$ theory both in the bulk and the boundary is negligible so that there is no deviation of the boundary theory from the conformal fixed point.  

{\bf Summary of the results}

We summarize below our findings of the present work as well as clarify on the notations pertaining to the work. We will be working in the regime $\ell\gg n\gg1$. 
By equating the contributions of the minimal twist operators as the stress tensor $T_{\m\n}$, current $J_\m$ and the singlet ($\e$) and the symmetric-traceless ($t_{ij}$) scalars , we find that the anomalous dimensions for the trace, symmetric-traceless and the antisymmetric-traceless operators in the large spin ($\ell\gg1$) sector are given by :
%\begin{itemize}
%\paragraph{Corrections to $\g_n$ from  $O^I_\ell$}
%\item
\begin{eqnarray}
\gamma_{n}^{I,\ell}&=& A_T\, \frac{P_T \, n^4}{\ell^{2}} +A_J\,(N-1)\frac{P_J\, n^2}  {\ell^{2}}+A_\e\,\frac{P_\e\, n^{2y_\e}}{\ell^{\D_\e}}+A_t\frac{(N^2+N-2)}{N} \frac{P_t \, n^{2y_t}}{\ell^{\D_t}}\, ,\nn
%\end{eqnarray}
%\item
%\paragraph{Corrections to $\g_n$ from $O^A_\ell$}
%\begin{eqnarray}
\gamma_{n}^{A,\ell}&=& A_T\, \frac{P_T \, n^4}{\ell^{2}}+A_J\, \frac{P_J \,n^2}{\ell^{2}}+ \frac{A_\e\,P_\e\, n^{2y_\e}}{\ell^{\D_\e}}-A_t\,\frac{(N+2)}{N}\frac{P_t\, n^{2y_t}}{\ell^{\D_t}}\, ,\nn
%\end{eqnarray}
%\item
%\paragraph{Corrections to $\g_n$ from  $O^S_\ell$}
%\begin{eqnarray}
\gamma_{n}^{S}&=&  A_T\,\frac{P_T\, n^4}{\ell^{2}}-A_J\,\frac{P_J\, n^2}{\ell^{2}}+A_\e\,\frac{P_\e\, n^{2 y_\e}}{\ell^{\D_\e}}+A_t\,\frac{(N-2)}{N} \frac{P_t\, n^{2y_t}}{\ell^{\D_t}} \, ,
\end{eqnarray}
%\end{itemize}

where
\begin{align}
\begin{split}
A_{T,\,J,\,\e,\,t}=-\frac{\G(2y+1)\G(2y+2) }{\G(1+y)^4\G(\D_\phi-\frac{\ta_m}{2})^2\G(\D_\f+y-1)^2}\G(\D_\phi)^2 \G(\D_\f-1)^2
\end{split}
\end{align}

and $y=\ell_m+\frac{\ta_m}{2}-1$, can take values  $2, 1, \frac{\D_\e-2}{2} $ and $\frac{\D_t-2}{2}$ for stress tensor, current, singlet scalar and symmetric tensor scalar respectively.

For the sample $O(2)$ model, the shifts in the binding energy from the bulk side due to the gravity and the gauge interactions are,
 \begin{equation}
   \delta E_{orb}=-\frac{16 \pi \, G_N}{\Omega_3}\frac{n^4}{\ell^2}\,,
 \end{equation}
and,
\begin{equation}
\d E_{orb}^J= \frac{\kappa^2 g^2}{2 \pi^2}\frac{n^2}{\ell^2}\,.
\end{equation}

which matches with the boundary calculation for the $O(2)$ models with a specific matching between the bulk and the boundary quantities. The above contribution due to the current is for a particular set of large spin operators. This sign flips for the other set of operators.

The remaining paper is organized as follows. In section \eqref{funda} we review the details of \cite{Li:2015rfa} as well as extend the analysis for the anomalous dimensions in the case of non zero twists $(n)$ considering separately the cases of the stress tensor, conserved current and the singlet and the symmetric-traceless scalars. Section \eqref{largen} describes the form of the anomalous dimensions for the large spin operators for the limit $\ell\gg n\gg1$ case. Note that the leading universal term in this case, depends on the twist $(n)$ and a certain combination of the minimal twist $\tau_m$ and the spin $\ell_m$ of the minimal twist operators. We plot the behaviour of the anomalous dimensions for the case when the minimal twist operator is either the stress tensor $T_{\m\n}$ or the conserved current $J_\m$. In section \eqref{epsilonexp}, we establish the known results in the literature about the large spin double twist operators in an $\e-$expansion from the conventional bootstrap. The results in this section are in complete agreement with the previously well known results in \cite{wilsonkogut} and so on. The next section \eqref{holo} describes the holographic counterpart of the calculations for the example case of the $O(2)$ model. We explain the details and the subtleties involved in the calculation and also point out the mapping between the corresponding quantities in the bulk and boundary theory.  We end the paper with a discussion on the possible future works and directions.

\section{$O(N)$ Fundamentals}\label{funda}
In this section we will use the conformal bootstrap for CFTs with an $O(N)$ symmetry. The details can be found in \cite{Rattazzi:2010yc, Vichi:2011ux, Poland:2011ey,  Kos:2013tga, Li:2015rfa}. We focus on theories containing a scalar field $\phi_i$ in the fundamental representation of $O(N)$ in $d=4$. Our goal is   to compute the anomalous dimension $\g(n,\ell)$ for the double-twist operators, defned  in \cite{Li:2015rfa} for non zero $n$. We begin by writing the four point correlation function $\langle \phi_{i_1}(x_1)\,\phi_{i_2}(x_2)\,\phi_{i_3}(x_3)\,\phi_{i_4}(x_4) \rangle$ in the s-channel and t-channel. The equality of s-channel and t-channel gives the bootstrap equation\cite{Li:2015rfa},
\begin{eqnarray}\label{bstpeq}
\left(\frac{u}{v}\right)^{\Delta_\phi} I_t (v,u) &=& \frac{1}{N} I_s (u,v)+\left(1-\frac{1}{N} \right) A_s (u,v)+ \left(1+\frac{1}{N}-\frac{2}{N^2}\right)S_s(u,v)\, ,\nn
\left(\frac{u}{v}\right)^{\Delta_\phi} A_t (v,u) &=& \frac{1}{2} I_s (u,v)+\frac{1}{2}  A_s (u,v)-\half \left(1+\frac{2}{N}\right)S_s(u,v)\, ,\nn
\left(\frac{u}{v}\right)^{\Delta_\phi} S_t (v,u) &=& \frac{1}{2} I_s (u,v)-\frac{1}{2}  A_s (u,v)+\half \left(1-\frac{2}{N}\right)S_s(u,v)\,.
\end{eqnarray}
We focus on the regime $u \ll v \ll 1$. In the (12)-(34) channel we have contributions from the identity operator, singlet scalars $\e$, symmetric tensor scalars $t_{ij}$, the current $J_\mu$ and the stress tensor $T_{\mu \nu}$. We assume the current and  stress tensor  to be conserved, so that they are at the unitarity bounds. In the limit $u \ll v \ll 1$ the s-channel blocks take the following form,

\begin{eqnarray}\label{rhs}
I_s(u,v) &\approx& 1+ P_\e g_{\D_\e,0} (u,v) + P_T g_{d-2,2} (u,v)\, ,\nn
A_s(u,v) &\approx& P_J g_{d-2,1} (u,v)\, ,\nn
S_s(u,v) &\approx& P_t g_{\D_t,0}(u,v)\,.
\end{eqnarray}

Here $g_{\tau,\ell}$ is a conformal block for an operator exchange of twist $\tau$ and spin $\ell$. In the (14)-(32) channel we have three types of double-twist operators:
\begin{equation}
O^I_{n,\ell} =\phi_i \square^n \partial^{\ell} \phi_i,~~~O^A_{n,\ell}=\phi_{[i} \square^n\partial^{\ell} \phi_{j]},~~~O^S_{n,\ell}=\phi_{(i} \square^n\partial^{\ell} \phi_{j)} -\frac{1}{N} \delta_{ij} \phi_k \square^n\partial^{\ell} \phi_k.
\end{equation}
The cross-channel conformal blocks are given by,
\begin{eqnarray}
I_t(v,u) &\approx& \sum_{\ell+} P_{O^I_\ell } g_{2 \D_\f+2n+\gamma^I}(v,u)\, ,\nn
A_t(v,u) &\approx& \sum_{\ell-} P_{O^A_\ell } g_{2 \D_\f+2n+\gamma^A}(v,u)\, ,\nn
S_t(v,u) &\approx& \sum_{\ell+} P_{O^S_\ell } g_{2 \D_\f+2n+\gamma^S}(v,u) \, .
\end{eqnarray}
Here the notation $\ell+$ and $\ell-$ means that the sum runs over even and odd spins respectively.
%{\color{red}{{Explanation for $\ell+$ and $\ell-$ ??}}}\nn
The leading contributions of \eqref{bstpeq} in the limit  $u \ll v \ll 1$ give,
\begin{eqnarray}
\frac{1}{N} &\approx&  \left(\frac{u}{v}\right)^{\Delta_\phi} I_t (v,u)\,,\nn
\half & \approx& \left(\frac{u}{v}\right)^{\Delta_\phi} A_t (v,u)\,, \nn
\half &\approx& \left(\frac{u}{v}\right)^{\Delta_\phi} S_t (v,u)\,.
\end{eqnarray}
As shown in \cite{Fitzpatrick:2012yx} if we write the cross-ratios as $u=z \bar{z}$ and $v=(1-z)(1-\bar{z})$ then at large $\ell$ in the 14-23 channel, the $\ell,z$ dependence of a conformal block separates from the $\tau,v$ dependence. Then we can write the above as,
\begin{eqnarray}\label{lo}
\frac{1}{N} &\approx&  \sum_{\ta} \bigg({\rm{lim}}_{z\rightarrow 0}\, z^{\D_\f} \sum_{\ell+} P_{O^I} k_{2 \ell} (1-z)\bigg) v^{\ta/2-\D_\f} (1-v)^{\D_\f} F^{(d)} (\ta,v),\nn
\frac{1}{2} &\approx&  \sum_{\ta} \bigg({\rm{lim}}_{z\rightarrow 0}\, z^{\D_\f} \sum_{\ell-} P_{O^A} k_{2 \ell} (1-z)\bigg) v^{\ta/2-\D_\f} (1-v)^{\D_\f} F^{(d)} (\ta,v),\nn
\frac{1}{2} &\approx&  \sum_{\ta} \bigg({\rm{lim}}_{z\rightarrow 0}\, z^{\D_\f} \sum_{\ell+} P_{O^S} k_{2 \ell} (1-z)\bigg) v^{\ta/2-\D_\f} (1-v)^{\D_\f} F^{(d)} (\ta,v).
\end{eqnarray}

Here $k_\b(x)={}_2F_1(\b/2,\b/2,\b,x)$. Since $F^{(d)} (\ta,v)$ \footnote{We will be working in $d=4$ . However  we can generalise this to genral $d$.}around small $v$ begins with a constant \cite{Fitzpatrick:2012yx,Kaviraj:2015xsa}, we have $\ta= 2\D_\f +2n$ in the spectrum. By matching the contribution of the LHS to the RHS of \eqref{lo}we get ,
\begin{eqnarray}
N P_{O^I_{n,\ell} } = P_{O^A_{n,\ell} } = P_{O^S_{n,\ell} } =P_{\D_\f, \D_\f}.
\end{eqnarray}
The MFT coefficients take the following form \cite{Fitzpatrick:2012yx},
\begin{equation}
P_{\D_\f, \D_\f}(n,\ell) = \frac{(1+(-1)^{\ell})\, {(\D_\f-1)^2_n\,(\D_\f)^2_{n+\ell}}}{\ell!\, n!\,(\ell+2)_n\, (2\D_\f+n-3)_n\, (2 \D_\f+2n+\ell-1)_\ell\, (2\D_\f+n+\ell-2)_n}\,,
\end{equation}
where the Pochhammer symbol $(a)_b=\G(a+b)/\G(a)$. In the large $\ell$ limit one can approximate
\begin{eqnarray}
P_{\D_\f, \D_\f} \overset{\ell \gg 1}\approx q_{\D_\f, n} \frac{\sqrt{\pi}}{2^{2 \D_\f+2n+2\ell}}\, \ell^{2\D_\f-3/2}\,,
\end{eqnarray}
with
\begin{eqnarray}
q_{\D_\f} =\frac{8}{\Gamma(\D_\f)^2} \frac{{(1-d/2+\D_\f)_n}^2}{n! (1-d+n+2 \D_\f)_n}\,.
\end{eqnarray}

Now we focus our attention on the subleading corrections to the bootstrap equation \eqref{bstpeq}.The subleading corrections are characterized by the anomalous dimension $\g(n, \ell)$ and the twist is given by $\tau(n, \ell)=2 \D_\f+2n+\g(n, \ell)$. We need to match the coefficients of the terms ${v^n\, logv}$ on both sides of \eqref{bstpeq} to find the corrections to the anomalous dimensions. One should refer to \cite{Fitzpatrick:2012yx, Komargodski:2012ek, Kaviraj:2015cxa} for the details. We will have four different contributions from the singlet scalars $\e$, symmetric tensor scalars $t_{ij}$, the current $J_\mu$ and the stress tensor $T_{\mu \nu}$ \eqref{rhs}. While computing $\g(n, \ell)$ we will frequently encounter the following the sums $A_i$ and $B_i$.

\begin{eqnarray}\label{sum}
A_i &&=\frac{1}{8} \Gamma\left(\Delta_\phi-\frac{\tau_m}{2}\right)^2 \sum_{\alpha=0}^{n} \gamma_{n-\alpha, i} \frac{q_{\Delta_\phi,{n-\alpha}} {(\frac{\tau}{2}-1)}_{n-\alpha}^2}{(\tau-2)_{n-\alpha} (n-\alpha)!}\,,\nn
 B_i &&= -\frac{P_i}{4}\frac{\Gamma(\tau_m+2 \ell_m)}{\Gamma(\frac{\tau_m}{2}+\ell_m)^2} {\frac{(\ell_m+\frac{\tau_m}{2})_n}{{(n!)}^2}}^2 \nn   && \times _3F_2\left[\{-n,-n,-1-\ell_m+\Delta_\phi-\frac{\tau_m}{2}\},\{1-\ell_m-n-\frac{\tau_m}{2}
 ,1-\ell_m-n-\frac{\tau_m}{2}
  \},1\right]\,,
 \end{eqnarray}
  where,
 \begin{equation}
 \gamma(n, \ell) =\frac{\gamma_{i,n}}{\ell^{\tau_m}}.
 \end{equation}
 and $i=T, J, \e, t $ for the stress-tensor, current, singlet scalar and symmetric tensor exchange respectively. in the $O(N)$ model, the bootstrap equation \eqref{bstpeq} is augmented by $N$-dependent factors as we write below for various cases.
%The $P_i$ in $B_i$ is replaced by $P_T$, $P_J$, $P_\e$ and $P_t$ for the stress-tensor, current, singlet scalar and symmetric tensor exchange respectively.

\subsection{Stress-tensor exchange}
For stress tensor exchange in $d=4$, $\tau_m=d-2=2$ and $\ell_m=2$ and $P=P_T$.  We have the following equations for  ${\gamma_{T,n}}^{I}$, ${\gamma_{T,n}}^{A}$ and ${\gamma_{T,n}}^{S}$ respectively:
 \begin{eqnarray}
 \frac{1}{N} A_T &=& \frac{1}{N} B_T\, , \nn %~~~~~\rm{gives~~ {\gamma_T}^{I,\ell}} \nn
 \half A_T &=& \half B_T \, , \nn%~~~~~~%\rm{gives ~~\gamma_T}^{A,\ell} \nn 
\half A_T &=& \half B_T\, . %~~~~~~%\rm{gives~~\gamma_T}^{S,\ell}.
\end{eqnarray}
For $n=0$, we have
% % % % % % % % % %
\begin{eqnarray}\label{strten}
\gamma_{T,0}^{I}&=&-\frac{P_T \Gamma \left(\Delta _{\phi }\right){}^2 \Gamma \left(2 l_m+\tau _m\right)}{4 \Gamma \left(l_m+\frac{\tau _m}{2}\right){}^2 \Gamma \left(\Delta _{\phi }-\frac{\tau _m}{2}\right){}^2}\, , \nn
\gamma_{T,0}^{A}&=&-\frac{P_T \Gamma \left(\Delta _{\phi }\right){}^2 \Gamma \left(2 l_m+\tau _m\right)}{4 \Gamma \left(l_m+\frac{\tau _m}{2}\right){}^2 \Gamma \left(\Delta _{\phi }-\frac{\tau _m}{2}\right){}^2}\, ,\nn
\gamma_{T,0}^{S}&=&-\frac{P_T \Gamma \left(\Delta _{\phi }\right){}^2 \Gamma \left(2 l_m+\tau _m\right)}{4 \Gamma \left(l_m+\frac{\tau _m}{2}\right){}^2 \Gamma \left(\Delta _{\phi }-\frac{\tau _m}{2}\right){}^2}\, .
\end{eqnarray}
% % % % % % % % % %
%\beq
%\gamma_{T,0}^{I,\ell}=-\frac{P_T}{4}\frac{\Gamma(\tau_m+2\ell_m)}{\Gamma(\frac{\tau_m}{2}+\ell_m)^2} \frac{{\Gamma(\D_\f)}^2}{{\Gamma(\D_\f-{\tau_m}/2)}^2}\nn
%\gamma_{T,0}^{A,\ell}=-\frac{P_T}{4}\frac{\Gamma(\tau_m+2\ell_m)}{\Gamma(\frac{\tau_m}{2}+\ell_m)^2} \frac{{\Gamma(\D_\f)}^2}{{\Gamma(\D_\f-{\tau_m}/2)}^2}\nn
%\gamma_{T,0}^{S,\ell}=-\frac{P_T}{4}\frac{\Gamma(\tau_m+2\ell_m)}{\Gamma(\frac{\tau_m}{2}+\ell_m)^2} \frac{{\Gamma(\D_\f)}^2}{{\Gamma(\D_\f-{\tau_m}/2)}^2}.
%\eeq
Thus the corrections due to stress-tensor exchange are negative. % and imply attractive gravitational interaction in the bulk theory.

 \subsection{Current exchange}
Here $\tau_m=d-2=2$ and $\ell_m=1$ and $P=P_J$.  We have the following equations for ${\gamma_J}^{I}$, ${\gamma_J}^{A}$ and ${\gamma_J}^{S}$ respectively:
 \begin{eqnarray}
 \frac{1}{N} A_J &=& \bigg(1-\frac{1}{N} \bigg) B_J\, ,\nn  %~~~~~\rm{gives~~ {\gamma_J}^{I,\ell}} \nn
\half A_J &=& \half B_J \, , \nn %~~~~~~\rm{gives ~~\gamma_J}^{A,\ell} \nn 
\half A_J &=& -\half B_J\, . %~~~~~~\rm{gives~~\gamma_J}^{S,\ell}.
\end{eqnarray}
For $n=0$, we have
\begin{eqnarray}\label{current}
\gamma_{J,0}^{I}&=&-(N-1) \frac{P_J}{4}\frac{\Gamma(\tau_m+2\ell_m)}{\Gamma(\frac{\tau_m}{2}+\ell_m)^2} \frac{{\Gamma(\D_\f)}^2}{{\Gamma(\D_\f-{\tau_m}/2)}^2}\, , \nn
\gamma_{J,0}^{A}&=&-\frac{P_J}{4}\frac{\Gamma(\tau_m+2\ell_m)}{\Gamma(\frac{\tau_m}{2}+\ell_m)^2} \frac{{\Gamma(\D_\f)}^2}{{\Gamma(\D_\f-{\tau_m}/2)}^2}\, , \nn
\gamma_{J,0}^{S}&=&\frac{P_J}{4}\frac{\Gamma(\tau_m+2\ell_m)}{\Gamma(\frac{\tau_m}{2}+\ell_m)^2} \frac{{\Gamma(\D_\f)}^2}{{\Gamma(\D_\f-{\tau_m}/2)}^2}\, .
\end{eqnarray}
The signs of anomalous dimensions depend on the representation of the operator.

\subsection{Singlet scalar exchange}
Here $\tau_m=\D_\e$ and $\ell_m=0$ and $P=P_\e$.  We have the following equations for ${\gamma_\e}^{I}$, ${\gamma_\e}^{A}$ and ${\gamma_\e}^{S}$ respectively:
 \begin{eqnarray}
 \frac{1}{N} A_\e &=& \frac{1}{N} B_\e\, ,\nn %~~~~~\rm{gives~~ {\gamma_\e}^{I,\ell}} \nn
 \half A_\e &=& \half B_\e \,, \nn %~~~~~~\rm{gives ~~\gamma_\e}^{A,\ell} \nn 
\half A_\e &=& \half B_\e\, . %~~~~~~\rm{gives~~\gamma_\e}^{S,\ell}.
\end{eqnarray}
For $n=0$,  we have
\begin{eqnarray}\label{singlet}
\gamma_{\e,0}^{I} &=&- \frac{P_\e}{4}\frac{\Gamma(\tau_m+2\ell_m)}{\Gamma(\frac{\tau_m}{2}+\ell_m)^2} \frac{{\Gamma(\D_\f)}^2}{{\Gamma(\D_\f-{\tau_m}/2)}^2}\, , \nn
\gamma_{\e,0}^{A}&=&-\frac{P_\e}{4}\frac{\Gamma(\tau_m+2\ell_m)}{\Gamma(\frac{\tau_m}{2}+\ell_m)^2} \frac{{\Gamma(\D_\f)}^2}{{\Gamma(\D_\f-{\tau_m}/2)}^2}\, ,\nn
\gamma_{\e,0}^{S}&=&-\frac{P_\e}{4}\frac{\Gamma(\tau_m+2\ell_m)}{\Gamma(\frac{\tau_m}{2}+\ell_m)^2} \frac{{\Gamma(\D_\f)}^2}{{\Gamma(\D_\f-{\tau_m}/2)}^2}\, .
\end{eqnarray}
\subsection{Symmetric tensor scalar exchange}
Here $\tau_m=\D_t$ and $\ell_m=0$ and $P=P_t$.  We have the following equations for ${\gamma_t}^{I}$, ${\gamma_t}^{A}$ and ${\gamma_t}^{S}$:
 \begin{eqnarray}
 \frac{1}{N} A_t &=& \bigg( 1+\frac{1}{N} -\frac{2}{N^2}\bigg) B_t \, , \nn %~~~~~\rm{gives~~ {\gamma_t}^{I,\ell}} \nn
 \half A_t &=& -\half \bigg(1+\frac{2}{N}\bigg) B_t \, , \nn  %~~~~~~\rm{gives %~~\gamma_t}^{A,\ell} \nn 
\half A_t &=& \half \bigg(1-\frac{2}{N}\bigg) B_t\, . %~~~~~~\rm{gives~~\gamma_t}^{S,\ell}.
\end{eqnarray}
For $n=0$, we have
\begin{eqnarray}\label{symmten}
\gamma_{t,0}^{I}&=&-\bigg(\frac{N^2+N-2}{N}\bigg) \frac{P_t}{4}\frac{\Gamma(\tau_m+2\ell_m)}{\Gamma(\frac{\tau_m}{2}+\ell_m)^2} \frac{{\Gamma(\D_\f)}^2}{{\Gamma(\D_\f-{\tau_m}/2)}^2} \, , \nn
\gamma_{t,0}^{A}&=&\bigg(\frac{N+2}{N}\bigg) \frac{P_t}{4}\frac{\Gamma(\tau_m+2\ell_m)}{\Gamma(\frac{\tau_m}{2}+\ell_m)^2} \frac{{\Gamma(\D_\f)}^2}{{\Gamma(\D_\f-{\tau_m}/2)}^2}\, , \nn
\gamma_{t,0}^{S}&=&-\bigg(\frac{N-2}{N}\bigg)\frac{P_t}{4}\frac{\Gamma(\tau_m+2\ell_m)}{\Gamma(\frac{\tau_m}{2}+\ell_m)^2} \frac{{\Gamma(\D_\f)}^2}{{\Gamma(\D_\f-{\tau_m}/2)}^2}\, .
\end{eqnarray}
Here also we have corrections of either sign.
Thus \eqref{strten}, \eqref{current}, \eqref{singlet}, \eqref{symmten} reproduces the results given in \cite{Li:2015rfa}.

\subsection{Pattern for Anomalous dimensions for $n \neq 0$}
Now we want to compute $\g_n$ for non zero $n$. We will consider the corrections due to stress tensor, current, singlet scalar and symmetric tensor exchange separately.
For stress-tensor we have,
\begin{eqnarray}
\gamma_{T,n}^{I, A, S}&=& \sum_{m=0}^{n} C_{n,m} {B^T}_m\, .
\end{eqnarray}
% % % % % % % % % % %For current % % % % % % % % % % %
%\subsubsection{For current}
For current,
\begin{eqnarray}
\gamma_{J,n}^{I}&=& (N-1)\sum_{m=0}^{n} C_{n,m} {B^J}_m \, , \nn
%\end{eqnarray}
%\begin{eqnarray}
\gamma_{J,n}^{A}&=& \sum_{m=0}^{n} C_{n,m} {B^J}_m \, ,\nn
%\end{eqnarray}
%\begin{eqnarray}
\gamma_{J,n}^{S}&=& -\sum_{m=0}^{n} C_{n,m} {B^J}_m\, .
\end{eqnarray}
%\subsubsection{For singlet scalar}
For singlet scalar, 
\begin{eqnarray}
\gamma_{\e,n}^{I, A, S}&=& \sum_{m=0}^{n} C_{n,m} {B^\e}_m \, .
\end{eqnarray}
For symmetric tensor,
\begin{eqnarray}
\gamma_{t,n}^{I}&=& \frac{(N^2+N-2)}{N}\sum_{m=0}^{n} C_{n,m} {B^t}_m \, , \nn
%\end{eqnarray}
%\begin{eqnarray}
\gamma_{t,n}^{A}&=&-\frac{(N+2)}{N} \sum_{m=0}^{n} C_{n,m} {B^t}_m \, ,\nn
%\end{eqnarray}
%\begin{eqnarray}
\gamma_{t,n}^{S}&=& \frac{(N-2)}{N}\sum_{m=0}^{n} C_{n,m} {B^t}_m\, .
\end{eqnarray}
where
\begin{eqnarray}
B^i_m &=& -\frac{P_i}{4}\frac{\Gamma(\tau_m+2 \ell_m)}{\Gamma(\frac{\tau_m}{2}+\ell_m)^2} {\frac{(\ell_m+\frac{\tau_m}{2})_n}{{(m!)}^2}}^2 \nn && \times _3F_2 \left[\{-m,-m,-1-\ell_m+\Delta_\phi-\frac{\tau_m}{2}\},\{1-\ell_m-m-\frac{\tau_m}{2}
,1-\ell_m-m-\frac{\tau_m}{2}
\},1\right]\, ,\nn
\end{eqnarray}
and
\begin{eqnarray}
C_{n,m} &=& (-1)^{m+n}\, \frac{\Gamma(\D_\f)^2}{{(\D_\f-1)_m}^2}\,\frac{n!}{(n-m)!}\,\frac{{(2\D_\f+n-3)}_m}{\Gamma(\D_\f-\ta_m/2)^2}\, .
\end{eqnarray}
It is evident that the corrections to the anomalous dimensions can have either sign depending on the nature of the double-twist operators in the spectrum and also on $N$. The corrections to the anomalous dimensions for different operators add up to the following,

%\paragraph{Corrections to the anomalous dimensions from  operators $O^I_\ell$}
\begin{eqnarray}
\gamma^{I}(n,\ell)&=& \sum_{m=0}^{n} C_{n,m} \bigg(\frac{P_T}{\ell^{d-2}}+(N-1)\frac{P_J}{\ell^{d-2}}+\frac{P_\e}{\ell^{\D_\e}}+\frac{(N^2+N-2)}{N} \frac{P_t}{\ell^{\D_t}}\bigg) B_m\, , \nn
%\end{eqnarray}
%\paragraph{Corrections to the anomalous dimensions from  operators $O^A_\ell$}
%\begin{eqnarray}
\gamma^{A}(n,\ell)&=& \sum_{m=0}^{n} C_{n,m} \bigg(\frac{P_T}{\ell^{d-2}}+\frac{P_J}{\ell^{d-2}}+ \frac{P_\e}{\ell^{\D_\e}}-\frac{(N+2)}{N}\frac{P_t}{\ell^{\D_t}}\bigg) B_m \, ,\nn
%\end{eqnarray}
%\paragraph{Corrections to the anomalous dimensions from $O^S_\ell$}
%\begin{eqnarray}
\gamma^{S}(n,\ell)&=& \sum_{m=0}^{n} C_{n,m} \bigg(\frac{P_T}{\ell^{d-2}}-\frac{P_J}{\ell^{d-2}}+\frac{P_\e}{\ell^{\D_\e}}+\frac{(N-2)}{N} \frac{P_t}{\ell^{\D_t}}\bigg) B_m\, .
\end{eqnarray}

\section{Leading $n$ dependence of anomalous dimensions}\label{largen}
In this section we want to extract the leading $n$ dependence of the coefficients of the anomalous dimensions for large $n$. In doing so we will follow \cite{Kaviraj:2015xsa}. $\g^i_n$ can be written as,
\begin{equation}
\g^i_n=\sum_{m=0}^n a^i_{n,m}\,,
\end{equation}
where,
\begin{align}\label{anm}
\begin{split}
a^i_{n,m}=-&\frac{P_i(-1)^{m+n}\G(n+1)\G(2\D_\f+n+m-3)\G(\D_\phi)^2\G(\D_\phi-1)^2\G(\ta_m+2\ell_m)\G(\ell_m+\ta_m/2+m)^2}{4\,\G(\D_\f-1+m)^2\G(\D_\f-\frac{\ta_m}{2})^2\G(n-m+1)\G(m+1)^2\G(\frac{\ta_m}{2}+\ell_m)^4\G(2\D_\f+n-3)}\\
&\times _3F_2\left[\{-m,-m,-1-\ell_m+\Delta_\phi-\frac{\tau_m}{2}\},\{1-\ell_m-m-\frac{\tau_m}{2}
 ,1-\ell_m-m-\frac{\tau_m}{2}\},1\right]\,.
\end{split}
\end{align}
We can write ${}_3F_2$  as, 
\begin{equation}
_3F_2\left[\{-m,-m,x+2-y\},\{-m-y,-m-y\},1\right]=\sum_{k=0}^m \frac{(-m)_k\ ^2 (x+2-y)_k}{(-m-y)_k\ ^2\ k!}\,,
\end{equation}
where $x=\D_\phi-4$ and $y=\ell_m+\frac{\ta_m}{2}-1$. Now $a^i_{n,m}$ can be written as,
\begin{align}\label{anm1}
\begin{split}
a^i_{n,m}=-&\frac{P_i(-1)^{m+n}\G(n+1)\G(2\D_\f+n+m-3)\G(\D_\phi)^2\G(\D_\phi-1)^2\G(\ta_m+2\ell_m)\G(\ell_m+\ta_m/2+m)^2}{4\,\G(\D_\f-1+m)^2\G(\D_\f-\frac{\ta_m}{2})^2\G(n-m+1)\G(m+1)^2\G(\frac{\ta_m}{2}+\ell_m)^4\G(2\D_\f+n-3)}\\
&\times \sum_{k=0}^m \frac{(-m)_k\ ^2 (x+2-y)_k}{(-m-y)_k\ ^2\ k!}\,.
\end{split}
\end{align}
We want to extract the large $m$ dependence inside the summation.
The large $m$ expansion takes the following form,
\begin{equation}
\sum_{k=0}^m\frac{(-m)_k\ ^2 (x+2-y)_k}{(-m-y)_k\ ^2\ k!}\overset{m\gg1}{\approx}\frac{\G(2y+1)\G(m+x+3-y)}{\G(m+1)\G(x+y+d-1)}+\cdots\,,
\end{equation}
where $\cdots$ are the subleading terms.
%This can be explicitly checked in Mathematica.
Thus to the leading order,
\begin{align}
\begin{split}
a^i_{n,m}\approx&-\frac{P_i(-1)^{m+n}\G(n+1)\G(2\D_\f+n+m-3)\G(\D_\phi)^2\G(\D_\phi-1)^2\G(\ta_m+2\ell_m)\G(\ell_m+\ta_m/2+m)^2}{4\,\G(\D_\f-1+m)^2\G(\D_\f-\frac{\ta_m}{2})^2\G(n-m+1)\G(m+1)^2\G(\frac{\ta_m}{2}+\ell_m)^4\G(2\D_\f+n-3)}\\
&\times\frac{\G(2y+1)\G(m+x+3-y)}{\G(m+1)\G(x+y+d-1)}\,\nn
\approx&-\frac{P_i(-1)^{m+n}\G(2y+1)\G(2y+2)\G(\D_\phi)^2\G(\D_\phi-1)^2}{4\,\G(y+1)^4 \G(\D_\f-\frac{\ta_m}{2})^2 \G(2\D_\f+n-3) \G(\D_\f+y-1)}\\
&\times\frac{\G(2\D_\f+m+n-3)}{4\,\G(m+\D_\f-1)} \times \frac{n!}{m! (n-m)!} \times \left[\frac{\G(y+m+1)^2 \G(m+\D_\f-y-1)}{\G(1+m)^2 \G(m+\D_\f-1)}\right].
\end{split}
\end{align}
The leading term inside the bracket is $m^y$. So the  coefficient $a^i_{n,m}$, to the leading order is given by,
\begin{align}
\begin{split}
a^i_{n,m}\approx&-\frac{P_i(-1)^{m+n}\G(2y+1)\G(2y+2)\G(\D_\phi)^2\G(\D_\phi-1)^2}{4\,\G(y+1)^4 \G(\D_\f-\frac{\ta_m}{2})^2 \G(2\D_\f+n-3) \G(\D_\f+y-1)}\\
&\times\frac{\G(2\D_\f+m+n-3)}{\G(m+\D_\f-1)} \times \frac{n! \, m^y}{m! (n-m)!} \,.
\end{split}
\end{align}
Using the reflection formula,
\begin{equation}
\G(m+\D_\f-1)\G(2-m-\D_\f)=(-1)^m\frac{\pi}{\sin( (\D_\f-1)\pi)}\,,
\end{equation}
the coefficients $a^i_{n,m}$ takes the following form,
\begin{align}
\begin{split}
a^i_{n,m}=-&P_i\,(-1)^{n}\frac{\sin((\D_\f-1)\pi)}{\pi}\frac{ n!\, \G(2y+1)\G(2y+2)\G(\D_\phi)^2 \G(\D_\f-1)^2}{4\,\G(1+y)^4\G(\D_\phi-\frac{\ta_m}{2})^2\G(2\D_\f+n-3) \G(\D_\f+y-1)}\\
&\frac{m^y}{m!(n-m)!}\G(2\D_\f+m+n-3)\G(2-m-\D_\f)\,.
\end{split}
\end{align}
We can now use the integral representation of the product of the Gamma functions to simplify it further,
\begin{equation}
\G(m+n+2\D_\f-3)\G(2-m-\D_\phi)=\int_0^\infty\int_0^\infty dx d{\tilde{y}}\ e^{-(x+{\tilde{y}})}x^{m+n-4+2\D_\phi}{\tilde{y}}^{1-m-\D_\phi}\,.
\end{equation}
Thus $\g^i_n$ can be written as,
\begin{align}
\begin{split}
\g^i_{n}=-&P_i\,(-1)^{n}\frac{\sin((\D_\f-1)\pi)}{\pi}\frac{ n!\,\G(2y+1)\G(2y+2)\G(\D_\phi)^2 \G(\D_\f-1)^2}{4\,\G(1+y)^4\G(\D_\phi-\frac{\ta_m}{2})^2\G(2\D_\f+n-3) \G(\D_\f+y-1)}\\
&\int_0^\infty \int_0^\infty dx d{\tilde{y}}\ e^{-(x+{\tilde{y}})}x^{n-4+2\D_\phi}{\tilde{y}}^{1-\D_\phi} \sum_{m=0}^{n} \bigg(\frac{x}{\tilde{y}}\bigg)^m \frac{m^y}{m!(n-m)!}\,.
\end{split}
\end{align}
{To perform the summation over $m$ we need information about $y$. Since $y=\ell_m+\frac{\ta_m}{2}-1$, it can take any value. However, we can perform the summation only when $y$ is integer or half integer using the techniques given in \cite{Kaviraj:2015xsa}. We can  do the summation numerically for any $y$.
\subsection{Integer y}
For integer $y$,
\begin{align}
\begin{split}
\g^i_{n}=-&P_i\,\frac{\G(2y+1)\G(2y+2)\G(\D_\phi)^2 \G(\D_\f-1)^2}{4\,\G(1+y)^4\G(\D_\phi-\frac{\ta_m}{2})^2\G(\D_\f+y-1)^2}\,\left[\frac{\G(n+1) \G(-3+n+y+2 \D_\f)}{\G(n+1-y) \G(2 \D_\f+n-3)}\right]\, .
\end{split}
\end{align}
To extract the leading $n$ dependence  we need to look at the leading $n$ term inside the last bracket. The leading term in $n$ in the last bracket is $n^{2y}$. Thus to the leading order in $n$, $\g^i_{n}$ is given by
\begin{align}
\begin{split}
\g^i_{n}=-&P_i\,\frac{\G(2y+1)\G(2y+2)\G(\D_\phi)^2 \G(\D_\f-1)^2}{4\,\G(1+y)^4\G(\D_\phi-\frac{\ta_m}{2})^2\G(\D_\f+y-1)^2}\, n^{2y}\, .
\end{split}
\end{align}
\subsection{Half integer y}
For half integer $y$,
\begin{align}
\begin{split}
\g^i_{n}=-&P_i\,\frac{\G(2y+1)\G(2y+2)\G(\D_\phi)^2 \G(\D_\f-1)^2}{4\,\G(1+y)^4\G(\D_\phi-\frac{\ta_m}{2})^2\G(\D_\f+y-1)^2}\,\left[\frac{ \G(-3+n+y+2 \D_\f)}{\G(2 \D_\f+n-3)}\right]\, n^y \, .
\end{split}
\end{align}
The term in the last bracket goes as $n^y$ in the large $n$ limit. Thus to the leading order in $n$, $\g^i_{n}$ is given by
\begin{align}
\begin{split}
\g^i_{n}=-&P_i\,\frac{\G(2y+1)\G(2y+2)\G(\D_\phi)^2 \G(\D_\f-1)^2}{4\,\G(1+y)^4\G(\D_\phi-\frac{\ta_m}{2})^2\G(\D_\f+y-1)^2}\, n^{2y}\, .
\end{split}
\end{align}
Thus for both integer and half integer $y$ we get the same result for the leading $n$ dependence for $\g^i_{n}$,
\begin{align}
\begin{split}
\g^i_{n}=-&P_i\,\frac{\G(2y+1)\G(2y+2)\G(\D_\phi)^2 \G(\D_\f-1)^2}{4\,\G(1+y)^4\G(\D_\phi-\frac{\ta_m}{2})^2\G(\D_\f+y-1)^2}\, n^{2y}\, ,
\end{split}
\end{align}
where $y= \ell_m+\frac{\ta_m}{2}-1$.

%\subsection{ $\g^{\ell}_{n}$ for O(N) models in d=4}
For $O(N)$ models in four dimensions $y=2, 1, \frac{\D_\e-2}{2}$ and $\frac{\D_t-2}{2} $ for stress tensor, current, singlet scalar and symmetric tensor exchange respectively. We will use $ y_\e$ and $ y_t$ for $\frac{\D_\e-2}{2}$ and $\frac{\D_t-2}{2}$ respectively.
%\begin{itemize}
%\item
%For stress-tensor exchange $y=2$.
%\item
%For current exchange $y=1$.
%\item
%For singlet scalar exchange $y=\frac{\D_\e-2}{2}=y_\e$
%\item
%For symmetric tensor exchange $y=\frac{\D_t-2}{2}=y_t$
Thus we have the following corrections to the anomalous dimensions  $\g_n$ for three types of double-twist operators $O^I_\ell$, $O^A_\ell$ and   $O^S_\ell$ in $O(N)$ respectively,
%\paragraph{Corrections to $\g_n$ from  $O^I_\ell$}
\begin{eqnarray}
\gamma^{I}(n,\ell)&=& A_T\, \frac{P_T \, n^4}{\ell^{2}} +A_J\,(N-1)\frac{P_J\, n^2}  {\ell^{2}}+A_\e\,\frac{P_\e\, n^{2y_\e}}{\ell^{\D_\e}}+A_t\frac{(N^2+N-2)}{N} \frac{P_t \, n^{2y_t}}{\ell^{\D_t}} \, ,\nn
%\end{eqnarray}
%\paragraph{Corrections to $\g_n$ from $O^A_\ell$}
%\begin{eqnarray}
\gamma^{A}(n,\ell)&=& A_T\, \frac{P_T \, n^4}{\ell^{2}}+A_J\, \frac{P_J \,n^2}{\ell^{2}}+ \frac{A_\e\,P_\e\, n^{2y_\e}}{\ell^{\D_\e}}-A_t\,\frac{(N+2)}{N}\frac{P_t\, n^{2y_t}}{\ell^{\D_t}} \, ,\nn
%\end{eqnarray}
%\paragraph{Corrections to $\g_n$ from  $O^S_\ell$}
%\begin{eqnarray}
\gamma^{S}(n,\ell)&=&  A_T\,\frac{P_T\, n^4}{\ell^{2}}-A_J\,\frac{P_J\, n^2}{\ell^{2}}+A_\e\,\frac{P_\e\, n^{2 y_\e}}{\ell^{\D_\e}}+A_t\,\frac{(N-2)}{N} \frac{P_t\, n^{2y_t}}{\ell^{\D_t}}\, ,
\end{eqnarray}
where
\begin{align}
\begin{split}
A_{T,\,J,\,\e,\,t}=-\frac{\G(2y+1)\G(2y+2) }{4\,\G(1+y)^4\G(\D_\phi-\frac{\ta_m}{2})^2\G(\D_\f+y-1)^2}\G(\D_\phi)^2 \G(\D_\f-1)^2 \, ,
\end{split}
\end{align}
for $y=2, 1, \frac{\D_\e-2}{2} $ and $\frac{\D_t-2}{2}$ respectively.

%In a compact notation we can write the corrections in the following manner,
%
%\begin{equation}
%\g(n, \ell)_i \approx A_i\, P_i\, \frac{n^{2\,\ell_m+\ta_m-2}}{\ell^{\ta_m}}
%\end{equation}
%where $\ell_m$ and $\ta_m$ are the spin and twist of the operators being exchanged.

Note that the signs of the corrections depend on the representation of the double-twist operator. It can have either sign and depends on $N$.

In fig. \ref{stressncurrent}, we show the plots for $\g_n$ for different values of $\D_\f$ for stress tensor and current exchange. For $n \gg 1$, the coincidence of the plots for different $\D_\f$ shows the universality of the leading $n$ dependence of $\g_n$.
%\end{itemize}
%\section{Graphs}

\begin{figure}[!htpb]
  \centering
      \begin{subfigure}{.4\textwidth}
        \centering
        \includegraphics[width=\textwidth]{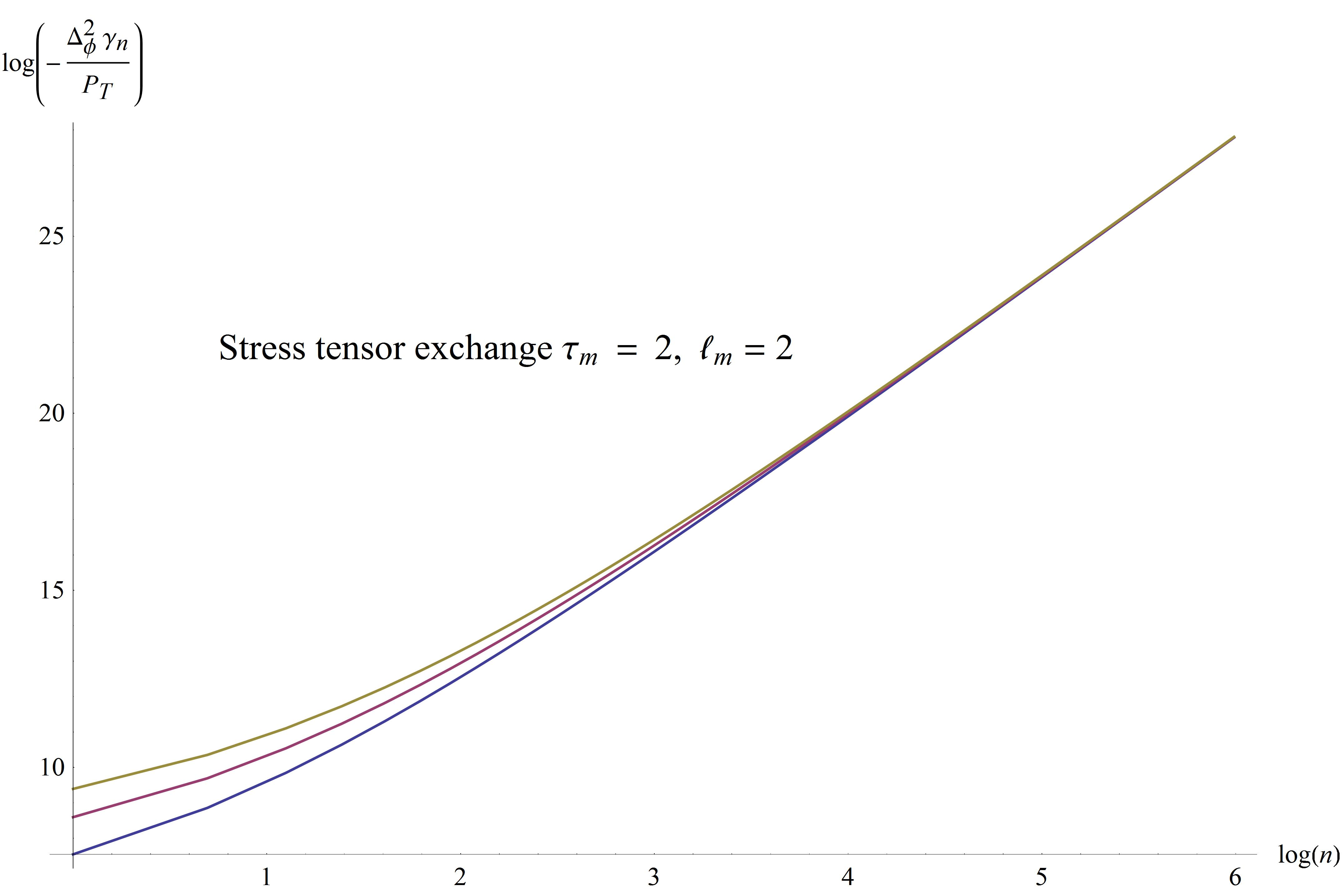}
        \caption{stress tensor exchange}
        \label{fig:stress}
      \end{subfigure}
       \begin{subfigure}{0.4\textwidth}
        \centering
        \includegraphics[width=\textwidth]{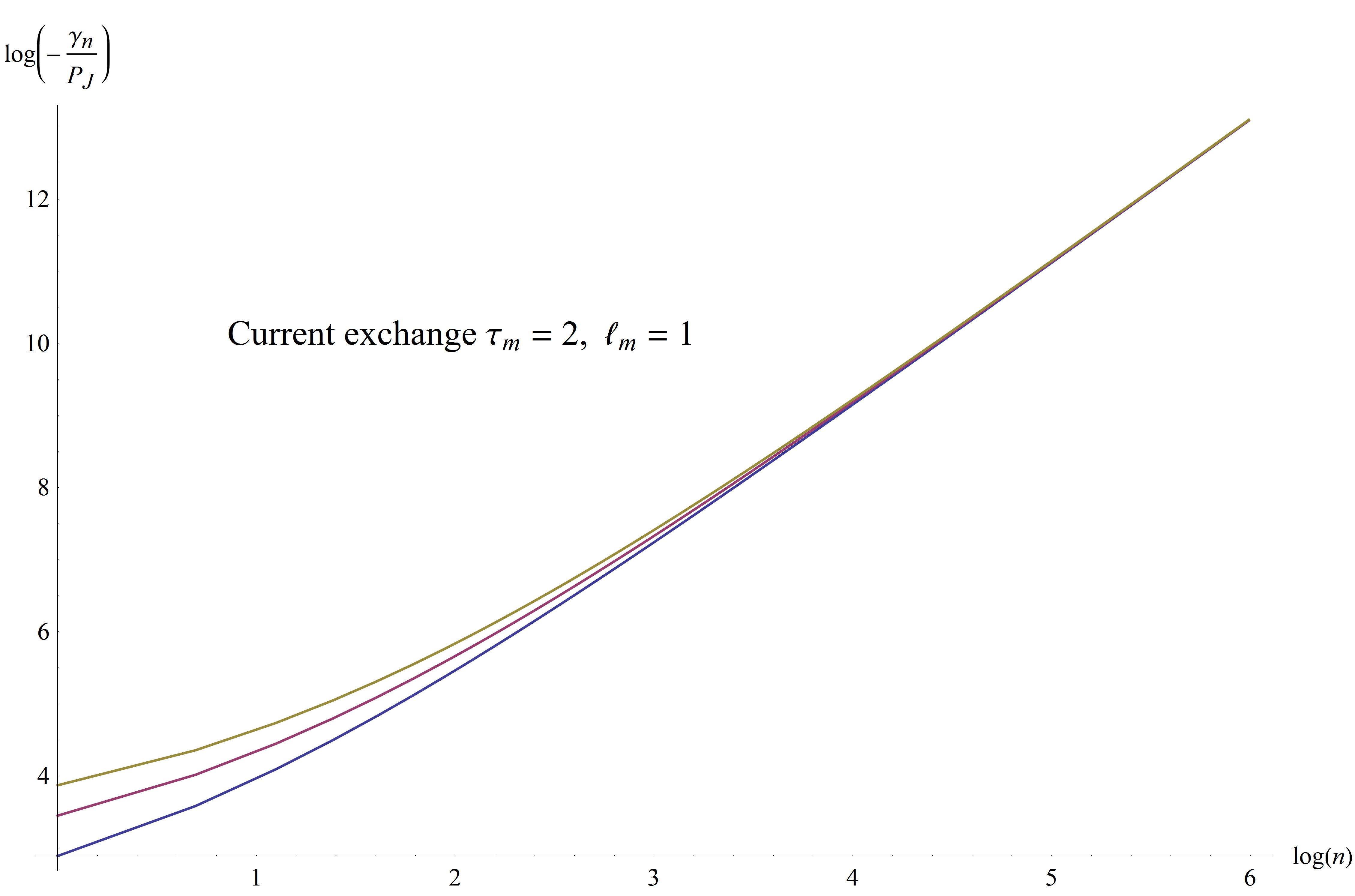}
        \caption{current exchange}
        \label{fig:current}
      \end{subfigure}\hfill%
    \caption{The figure shows the variation of the $\log (-\gamma_n)$ with $\log n$ for the current and the stress tensor exchange for different $\Delta_\f$. The normalizations $P_T$ and $P_J$ are for each of the current and stress tensor exchange.}
    \label{stressncurrent}
\end{figure}

\section{$\e$-expansion from Bootstrap}\label{epsilonexp}

In this section we demonstrate how the bootstrap analysis reproduces known results of double-twist operators. We will consider operators of the type,
\begin{equation}\label{operatorsinphi4}
	O^I_{\ell}=\phi_i \partial^\ell\phi_i\,, \hspace{0.5cm}O^A_{\ell}=\phi_{[i} \partial^\ell\phi_{j]}\,, \hspace{0.5cm}O^S_{\ell}=\phi_{(i} \partial^\ell\phi_{j)}-\frac{\delta_{ij}}{N}\phi_k \partial^\ell \phi_k\,,
\end{equation}
in the $\lambda (\phi_i \phi_i)^2$ theory in $d=4-\epsilon$ dimension. We will analyse the effect of two scalars, the singlet $O_1=\phi_i \phi_i$ and the symmetric and traceless $O_2=\phi_{(i}\phi_{j)}$ in the $s$-channel, on the above operators. Let us call their twists $\ta_1$ and $\ta_2$ respectively. Now, as shown in \cite{Li:2015rfa}, the anomalous dimensions of the above operators, due to the singlet scalar, is given by,
\begin{equation}\label{scalar1cont}
	\delta \gamma_\ell^I=\delta \gamma_\ell^A=\delta \gamma_\ell^S=- \frac{P_{1}\gamma_0^{\tau_{1},0}}{\ell^{\tau_{1}}}\,,
\end{equation}
and due to the symmetric traceless scalar, is given by,
\begin{equation}\label{scalar2cont}
	\frac{N}{N^2+N-2}\delta \gamma_\ell^I=-\frac{N}{N+2}\delta \gamma_\ell^A=\frac{N}{N-2}\delta \gamma_\ell^S=- \frac{P_{2}\gamma_0^{\tau_{2},0}}{\ell^{\tau_{2}}}\,.
\end{equation}
In the above, $\gamma_0^{\tau,\ell}$ is given by,
\begin{equation}\label{gamma0}
	\gamma_0^{\tau,\ell}=\frac{2 \Gamma (2 \ell +\tau ) \Gamma ^2\left(\Delta _{\phi }\right)}{\Gamma ^2\left(\ell +\frac{\tau }{2}\right) \Gamma ^2\left(-\frac{\tau }{2}+\Delta _{\phi }\right)}\,,
\end{equation}
with $\ell=0$ for scalar exchanges, and $\tau$ taking the values $\tau_{1}$ or $\tau_{2}$ according to the singlet or symmetric traceless exchange.

$P_1$ and $P_2$, the ope coefficients for the operators $O_1$ and $O_2$,  are also known (see \cite{Li:2015rfa}, \cite{polyakov}).  They are given by,
\begin{equation}
	P_1=\frac{2}{n}\hspace{0.5cm}\text{and}\hspace{0.5cm}P_2=1\,.
\end{equation}
Now the dimension of $\phi_i$ is $\Delta_\phi=(d-2)/2+O(\epsilon^2)$ and twist of the singlet scalar is given by,
\begin{equation}
	\tau_1 =(d-2)+\frac{(2+n) \epsilon }{8+n}+O\left(\epsilon ^2\right)\,,
\end{equation}
and that of the traceless symmetric scalar is,
\begin{equation}
	\tau _2=(d-2)+\frac{2 \epsilon }{8+n}+O\left(\epsilon ^2\right)\,.
\end{equation}
Using the above in \eqref{gamma0} and evaluating the anomalous dimensions of the operators \eqref{operatorsinphi4}, we get,
\begin{align}
	\label{gammaI}&\gamma_\ell^I=- \frac{P_{1}\gamma_0^{\tau_{1},0}}{\ell^{\tau_{1}}}-\left(\frac{N^2+N-2}{N}\right)\frac{P_{2}\gamma_0^{\tau_{2},0}}{\ell^{\tau_{2}}} \ \ = \ \ -\frac{3 (2+N) }{(8+N)^2}\frac{\epsilon ^2}{\ell^2}\,,\\\label{gammaA}&\gamma_\ell^A=- \frac{P_{1}\gamma_0^{\tau_{1},0}}{\ell^{\tau_{1}}}-\left(-\frac{N+2}{N}\right)\frac{P_{2}\gamma_0^{\tau_{2},0}}{\ell^{\tau_{2}}} \ \ = \ \ -\frac{(2+N) }{(8+N)^2 }\frac{\epsilon^2}{\ell^2}\,,\\\label{gammaS}&\gamma_\ell^S=- \frac{P_{1}\gamma_0^{\tau_{1},0}}{\ell^{\tau_{1}}}-\left(\frac{N-2}{N}\right)\frac{P_{2}\gamma_0^{\tau_{2},0}}{\ell^{\tau_{2}}} \ \ = \ \ -\frac{(6+N) }{(8+N)^2 }\frac{\epsilon ^2}{\ell ^2}\,.
\end{align}
Higher spin exchanges of minimal twists in the $s$-channel should also contribute to the above results. However if we assume the anomalous dimensions of such operators to start from $O(\epsilon^2)$, their effects show up at an higher order of $\epsilon$. So we can neglect them in our analysis.

It was shown in \cite{wilsonkogut} using standard feynman diagrams that the anomalous dimensions of $O_\ell^I$ and $O_\ell^S$ kinds of operators are given by,
\begin{align}
	&\gamma_{O^I_\ell}=\frac{N+2}{2(N+8)^2}\epsilon^2\left(1-\frac{6}{\ell(\ell+1)}\right)\,,\\ \text{and} \hspace{1cm}& \gamma_{O^S_\ell}=\frac{N+2}{2(N+8)^2}\epsilon^2\left(1-\frac{2(N+6)}{(N+2)\ell(\ell+1)}\right)\,.
\end{align}
In the above the 1-s inside the parentheses come from the anomalous dimensions of $\phi_i$. The anomalous dimensions we computed above in \eqref{gammaI}, \eqref{gammaA} and \eqref{gammaS} are only the spin dependent parts of the total anomalous dimensions, and for $\ell \gg 1$. Hence they agree very nicely with the above known reults. One should be able to incorporate the effect of other exchange operators systematically to go to the next order in $\epsilon$. It will also be interesting to use the techniques of \cite{Rychkov:2015naa} or \cite{Sen:2015doa} to reproduce these results.
 
 \section{Holographic calculation: Example O(2) model}\label{holo}
 In this section we will try to compare the double twist anomalous dimensions for an  $O(2)$ model in a holographic picture. For holography, we will implicitly assume that there is a large $N$ gauge theory with an Einstein gravity in the bulk. So essentialy we have a CFT with a large symmetry group that has a bulk dual, and also having two equi-dimensional scalars. This picture is similar to \cite{Fitzpatrick:2014vua}, except we have two scalars instead of one. We will match the anomalous dimensions for the $O(2)$ model in the field theory with this picture on the holographic side. We will consider an external charged scalar in the probe limit coupled with the Einstein action and the $U(1)$ gauge field. Then we are considering an $O(2)$ model as a probe in the field theory itself so that there is no significant deformation of the CFT. The bulk is just the low energy Einstein gravity with a charged scalar so that the zeroth order part of the dual is still $AdS_5$. 
 
With this bulk we add a charged complex scalar field $\phi$  coupled to gravity and gauge field \cite{Fitzpatrick:2011hh}:
 \begin{equation}\label{sphi}
 S=\frac{1}{\kappa^2}\int d^5 x \sqrt{-g} \big[R+6-\frac{1}{4 g^2}\, F^2-(D^{\mu}\phi)^\dagger (D_\mu \phi)-m^2 \phi^\dagger \phi\big]\,,
 \end{equation}
 where $D_\mu=\partial_\mu -i A_\mu$ and  $\kappa=\sqrt{8 \pi G_N}$ .% (The $U(1)$ charge is given by $q=\frac{2}{3} \D_\f$ where $\D_\f$ is the scaling dimension of the scalar field.
 
Note that we have redefined the gauge coupling $g$ and absorbed the charge of the scalar in the coupling so that there is no net charge appearing anywhere in the above action. Henceforth $g$ is our coupling. Our goal is to compute the leading order binding energies of generalized free fields in the bulk with large angular momentum, due to gravitational and gauge interactions in the bulk. According to \cite{Fitzpatrick:2011hh}, the gauge and graviton exchange deform the Hamiltonian as,
 \begin{equation}
 H_{free}\rightarrow H_{free} +  \delta H.
 \end{equation}
The $\d H$ is obtained for \eqref{sphi} by expanding in the interactions of the scalar $\phi$ with the gauge and gravity parts. We write the $\e H$ in terms of the interaction potential given by \cite{Fitzpatrick:2011hh},
\begin{equation}
\d H=V_{eff}[\phi,\phi^\dagger]=V_{quar}[\phi,\phi^\dagger]+\frac{\kappa}{2}A^\m J_\m[\phi,\phi^\dagger]+\frac{\kappa}{4}h^{\m\n}T_{\m\n}[\phi,\phi^\dagger]\,,
\end{equation}
where $V_{quar}[\phi,\phi^\dagger]$ is the quartic scalar interaction. The first order energy shift is given by the expectation value of the interaction Hamiltonian using the unperturbed wavefunction for the orbiting object. 
 Then the shift in energy is given by
 \begin{equation}
 \delta E =   \langle {n, \ell_{orb}} \mid \delta H \mid {n, \ell_{orb}}\rangle \nn\,.
 \end{equation}
 %First we want to compute the interaction due to the presence of a point mass and then we will evaluate the expectation value.
 
 To begin, let us consider the free theory $(\kappa\rightarrow0)$. We take $AdS_5$ metric in global coordinates,
 \begin{equation}
 ds^2= \frac{1}{\cos^2\rho} (dt^2-d\rho^2-\sin^2 \rho\, d\Omega_3^2)\,.
 \end{equation}
 We will work in units of  AdS radius $R_{AdS}=1$.  We now consider a free massive scalar field $\psi(x)$  in the bulk satisfying $(\nabla^2-m^2) \psi=0$. The wavefunction is given by,
 \begin{eqnarray}\label{wvfn}
 \psi_{n\,\ell\,J}(t, \rho, \Omega)&=&\frac{1}{N_{\Delta, n, \ell}} \, e^{-i E_{n, \ell}\,t}\, Y_{\ell, J}(\Omega)\, \bigg[\rm{sin}^{\ell} \rho \, \rm{cos}^\Delta \rho \, _2F_1 \bigg(-n, \Delta_\f+\ell+n, \ell+2, \rm{sin}^2 \rho\bigg)\bigg]\, ,\nn
  E_{n,\ell}&=& \Delta_\f+2n+\ell \, ,\nn
  m^2&=& \D_\f(\D_\f-4)\, ,
 \end{eqnarray}
 with normalizations
 \begin{equation}
  N_{\D, n, \ell} =(-1)^{\ell}\, \sqrt{\frac{n!\, \Gamma(\ell+2)\, \Gamma(\Delta_\f+n-1)}{\Gamma(n+\ell+2)\, \Gamma(\Delta_\f+n+\ell)}}\, , %\,\, \rm{with} \,\,\,\,d=4\,\, and\,\, r=\rm{tan}\rho
  \end{equation}
 where $Y_{\ell, J}(\Omega)$ are the normalised eigenstates of the Laplacian on $S^3$. Here the quantum numbers $n$ and $\ell$ denote the twist and angular momentum respectively.
 
The shift in energy due to gravitational and gauge interactions between the scalar fields in $AdS_5$ as an expansion in inverse distance corresponds to the CFT computation of anomalous dimensions of the large spin ($\ell \gg 1$) double-twist operators.

 Computing the first order energy shift due to gravitational interactions in the bulk is equivalent to computing the gravitational interaction of a scalar field in AdS-Schwarzschild black hole \cite{Fitzpatrick:2014vua}.
% To compute the first order energy shift due to gravitational interactions in the bulk,
 We start with the AdS-Schwarzschild{\footnote{We can replace the AdS-Schwarzschild black hole with the RN-AdS black hole. But this will give subleading corrections to the anomalous dimensions.}} black hole in five dimensions,
 \begin{equation}
 ds^2 = f(r)\, dt^2 -\frac{dr^2}{f(r)}- r^2 d\Omega_3^2\,,
 \end{equation}
 where, 
 \begin{equation}
 f(r)= 1-\frac{2M}{r^2}+r^2\,.
 \end{equation}
 and the mass of the black hole is 
 \begin{equation}
 M_{BH}= \frac{3\, \Omega_3\, M}{8 \pi \, G_N}\, .
 \end{equation}
 The shift in  energy to first order in $M$ is given by,
 \begin{eqnarray}
 \delta E_{orb}&=& \langle {n, \ell_{orb}} \mid\delta H\mid {n, \ell_{orb}}\rangle
  \nn &=& -M\,\int dr\, d\Omega_3 \, r^3 \, \langle {n, \ell_{orb}} \mid \frac{1}{r^2(1+r^2)^2} (\partial_t \phi)^2 + \frac{1}{r^2} (\partial_r \phi)^2 \mid {n, \ell_{orb}}\rangle\,,
 \end{eqnarray}
 where $r= \tan\rho$. Here the label `$orb$' implies that we are considering one mass, described by the scalar field, orbiting a second mass $M_{BH} $ at the origin of AdS, with relative angular momentum $\ell_{orb}$. We use the wavefunctions from\eqref{wvfn} to compute $\delta E_{orb} $ as,
  \begin{eqnarray}
\delta E_{orb}&=& -M \frac{\Gamma (\Delta_\f +1) \,(\ell_{orb}+2 n)^2\, \Gamma (\ell_{orb}+n+2)}{2 \Gamma (\ell_{orb}+2)\, \Gamma (n+\Delta_\f -1)} \nn &&
 \times \sum _{k=0}^n \frac{(-1)^k \Gamma (k+\ell_{orb}+1) \Gamma (k+\ell_{orb}+n+\Delta_\f ) }{k!\, \Gamma (k+\ell_{orb}+2) \,\Gamma (-k+n+1)\, \Gamma (k+\ell_{orb}+\Delta_\f +2)}\nn
 && \times \, _3F_2(k+\ell_{orb}+1, -n, \ell_{orb}+n+\Delta_\f ;\ell_{orb}+2,k+\ell_{orb}+\Delta_\f +2;1)\, .
  \end{eqnarray}
 For $n=0$, 
 \begin{eqnarray}
 \delta E_{orb} (n=0)&=& -\frac{4 \pi G_N\, M_{BH}}{3 \Omega_3} \D_\f (\D_\f-1)\frac{1}{\ell_{orb}}\,.
 \end{eqnarray}
 We can calculate $ \delta E_{orb}$ for $n=0, 1, 2, \cdots$ and get a general $n$ dependence. The leading $n$ dependence of the energy shift, in agreement with \cite{Kaviraj:2015xsa}  becomes,
 \begin{eqnarray}
  \delta E_{orb}&&= -2M \frac{1}{\ell_{orb}}\, \frac{\G(4)}{\G(2)\, \G(3)}\, n^2+ \cdots\nn
  &&=- 6\,M \frac{1}{\ell_{orb}}\, n^2 +\cdots \nn
  &&=- \frac{16\,\pi\, G_N\,M_{BH}}{\Omega_3} \frac{1}{\ell_{orb}}\, n^2 +\cdots\,.
 \end{eqnarray}
 As shown in \cite{Fitzpatrick:2014vua}, \cite{Kaviraj:2015cxa} this system is equivalent to two scalar objects rotating around the centre of AdS, with total angular momentum $\ell$.
 The relation between $\ell_{orb}$ and $\ell$ is given by $\ell_{orb} \approx \ell^2/n$ and $M_{BH} \approx n$ for large $n$. Thus we get
 \begin{equation}
   \delta E_{orb}=-\frac{16 \pi \, G_N}{\Omega_3}\frac{n^4}{\ell^2}\, .
 \end{equation}

 Finally, let us evaluate the shift in energy due to gauge interactions in the bulk following \cite{Fitzpatrick:2011hh}. We have considered one type of operator for which the current contribution is of a particular sign. In principle we can also consider the other set of large spin operators (the antisymmetric ones) for which this contribution comes with a negative sign. This also concurs for the two different signs of the contribution due to the conserved current for the symmetric traceless and antisymmetric large spin operators in the CFT.
 
 \begin{equation}
 \delta E^J_{orb} = \int dr \, d\Omega_3 \, r^3 \,\langle {n, \ell_{orb}}\mid J_0\, A^0\mid {n, \ell_{orb}}\rangle\,,
 \end{equation}
 where 
 \begin{eqnarray}
 J_\mu &=& ig\,(\phi \partial_\mu \phi^{\dagger}-\phi^{\dagger}\partial_{\mu}\phi)\, , \nn A^0 &=&-\frac{N_\D^2\, g}{2\,(\D_\f-1)} \bigg(\frac{1}{r^2(1+r^2)}-\frac{1}{r^2(1+r^2)^{\Delta_\f}}\bigg)\,,
 \end{eqnarray}
and
 \begin{equation}
 N_\D = \sqrt{\frac{\D_\f-1}{2 \pi^2}}\, 
 \end{equation}
in a gauge where the only surviving component of $A_{\mu}$ is $A_0$ as given in \cite{Fitzpatrick:2011hh}. 
%\footnote {We can choose a gauge where the only surviving component of $A_{\mu}$ is $A_0$.}

 Using the wavefunction, we find
 \begin{eqnarray}
 \delta E^J_{orb} &=& \frac{\kappa^2\,g^2}{2 \pi^2}\sum_{k,\alpha=0}^{n}\frac{E_{n,\ell_{orb}}}{N_{\D, n, \ell}^2}\frac{(-1)^{k+\alpha}\,\Gamma(n+1)^2\, (\Delta_\f+\ell_{orb}+n)_k\, (\Delta_\f+\ell_{orb}+n)_{\alpha}}{\Gamma(k+1)\, \Gamma(n-k+1)\,\Gamma(\alpha+1)\, \Gamma(n-\alpha+1)\,(\ell_{orb}+2)_k\, (\ell_{orb}+2)_\alpha}\nn && \times \int dr \frac{r^{3+2\ell_{orb}}}{(1+r^2)^{\Delta_\f+\ell}}\, \frac{r^2}{(1+r^2)^{k+\alpha}}\,\bigg[\frac{1}{r^2(1+r^2)}-\frac{1}{r^2(1+r^2)^{\Delta_\f}}\bigg]\,.
 \end{eqnarray}
 The $r$ integral gives,
 \begin{eqnarray}
 \int dr \frac{r^{3+2\ell_{orb}}}{(1+r^2)^{\Delta_\f+\ell_{orb}}}\, \frac{r^2}{(1+r^2)^{k+\alpha}}\,\bigg[\frac{1}{r^2(1+r^2)}-\frac{1}{r^2(1+r^2)^{\Delta_\f}}\bigg]&=& I_1-I_2\,,
  \end{eqnarray}
  where
  \begin{equation}
  I_1=\frac{\Gamma(1+k+\ell_{orb}+\alpha)\,\Gamma(\Delta_\f)}{2\, \Gamma(1+k+\ell_{orb}+\alpha+\Delta_\f)}\,.
  \end{equation}
  and
  \begin{equation}
    I_2=\frac{\Gamma(1+k+\ell_{orb}+\alpha)\,\Gamma(2\Delta_\f-1)}{2\, \Gamma(k+\ell_{orb}+\alpha+2\Delta_\f)}\,.
    \end{equation}
    Lets consider the contribution from $I_1$.
    
 Performing the first sum over $\alpha$ we get
 \begin{eqnarray}
&&\frac{\kappa^2\,g^2}{2 \pi^2} \sum_{k=0}^{\alpha} \frac{E_{n,\ell_{orb}}}{N_{\D, n, \ell}^2} \frac{(-1)^k\,\Gamma(n+1)\, (\Delta_\f+\ell_{orb}+n)_k}{\Gamma(k+1)\,\Gamma(n-k+1)\,(\ell+2)_k}\,\bigg[\frac{\Gamma(k+\ell_{orb}+1)\,\Gamma(\Delta_\f)}{2\Gamma(k+\ell_{orb}+\Delta_\f+1)}\bigg] \nn
&& \times \sum_{\alpha=0}^{n}\frac{(-1)^\alpha\, n! \,(\Delta_\f+\ell_{orb}+n)_\alpha\, (k+\ell_{orb}+1)_\alpha}{\alpha!\,(\ell_{orb}+2)_\alpha\,(n-\alpha)!\, (k+\ell_{orb}+\Delta_\f+1)_\alpha}\nn
 &&= \frac{\kappa^2\,g^2}{2 \pi^2}\sum_{k=0}^{\alpha} \frac{E_{n,\ell_{orb}}}{N_{\D, n, \ell}^2} \frac{(-1)^k\,\Gamma(n+1)\, (\Delta_\f+\ell_{orb}+n)_k}{\Gamma(k+1)\,\Gamma(n-k+1)\,(\ell_{orb}+2)_k}\,\bigg[\frac{\Gamma(k+\ell_{orb}+1)\,\Gamma(\Delta_\f)}{2\Gamma(k+\ell_{orb}+\Delta_\f+1)}\bigg]\nn &&\times _3F_2[\{1+k+\ell_{orb}, -n, \ell_{orb}+n+\Delta_\f\},\{2+\ell_{orb}, 1+k+\ell_{orb}+\Delta_\f\},1]\nn
 && =\frac{\kappa^2g^2}{2 \pi^2}\frac{1}{2\ell_{orb}}\,(\Delta_\f+2n-1)\,.
 \end{eqnarray} 
 
 Now we calculate the contributions coming from $I_2$. The first sum over $\alpha$ gives 
 \begin{eqnarray}
 &&\frac{\kappa^2\,g^2}{2 \pi^2}\,\sum_{k=0}^{\alpha} \frac{E_{n,\ell_{orb}}}{N_{n, \ell}^2} \frac{(-1)^k\,\Gamma(n+1)\, (\Delta_\f+\ell_{orb}+n)_k}{\Gamma(k+1)\,\Gamma(n-k+1)\,(\ell_{orb}+2)_k}\,\bigg[\frac{\Gamma(k+\ell_{orb}+1)\,\Gamma(2\Delta_\f-1)}{2\Gamma(k+\ell_{orb}+2\Delta_\f)}\bigg]\nn && \times \sum_{\alpha=0}^{n}\frac{(-1)^\alpha\, n! \,(\Delta_\f+\ell_{orb}+n)_\alpha\, (k+\ell_{orb}+1)_\alpha}{\alpha!\,(\ell_{orb}+2)_\alpha\,(n-\alpha)!\, (k+\ell_{orb}+2{\Delta_\f})_\alpha}\nn
 &&=\frac{\kappa^2\,g^2}{2 \pi^2} \sum_{k=0}^{\alpha} \frac{E_{n,\ell_{orb}}}{N_{n, \ell}^2} \frac{(-1)^k\,\Gamma(n+1)\, (\Delta_\f+\ell_{orb}+n)_k}{\Gamma(k+1)\,\Gamma(n-k+1)\,(\ell_{orb}+2)_k}\,\bigg[\frac{\Gamma(k+\ell_{orb}+1)\,\Gamma(2\Delta_\f-1)}{2\Gamma(k+\ell_{orb}+2\Delta_\f)}\bigg]\nn && \times _3F_2[\{1+k+\ell_{orb},-n,\ell_{orb}+n+\Delta_\f\},\{2+\ell_{orb}, k+\ell_{orb}+2\Delta_\f\},1]\nn
 && \approx \frac{1}{\ell_{orb}^{\Delta_\f}}.
 \end{eqnarray}
 Thus the contributions from $I_2$ are at a much higher order $O(1/\ell_{orb}^{\Delta_\f})$ and hence do not affect the leading order result.
 
 To the leading order in $\ell_{orb}$, the shift in energy due to gauge interactions is given by
 \begin{eqnarray}
 \delta E^J_{orb} &&=\frac{\kappa^2\,g^2}{2 \pi^2}\frac{1}{2\ell_{orb}}\,(\Delta_\f+2n-1) +\cdots\nn
 &&=\frac{\kappa^2\,g^2}{2 \pi^2}\frac{1}{\ell_{orb}}\,n+ \cdots\nn
 && \approx \frac{\kappa^2\,g^2}{2 \pi^2}\frac{n^2}{\ell^2}\,.
 \end{eqnarray} 
From the CFT bootstrap result we have the following predictions for the anomalous dimensions due to stress tensor and current exchange:
 \begin{eqnarray}\label{match}
 \gamma_{T}^{\ell}&=& -\frac{40 }{\pi^4 C_T} \, \frac{n^4}{\ell^{2}} \, ,\nn
  \gamma_{J}^{\ell}&=&  (-1)^O\, \frac{3 }{\pi^4\, C_J} \frac{ n^2}  {\ell^{2}} \, .
 \end{eqnarray}
 The sign of $\gamma_{J}^{\ell}$ depends on the nature of the double-twist operators. It is negative for $O^I$, $O^A$ and positive for $O^S$.
 
In four dimensions, we have used the relations  $S_d= 2 \pi^2=\Omega_{d-1}, P_T=\frac{8 \D_\f^2}{9 \pi^4 C_T},  P_J = \frac{4}{C_J}$,  $g^2\, \kappa^2=\frac{6}{\pi^2\, C_J}$ and $G_N=\frac{5}{C_T\, \pi^3}$ which reproduces \eqref{match}. This choice of the normalization is consistent with the results of \cite{Li2} and \cite{Fitzpatrick:2011hh}. 
 
To summarize our findings in this section, we have considered a specific example of a large $N$ CFT dual to an Einstein gravity residing on $AdS_5$, and the $O(2)$ model acting as a perturbation to this CFT. In the dual gravity the O(2) perturbation corresponds to a charged scalar field coupled to a U(1) gauge field. Hence the gravitational and gauge interactions, computed from the respective energy shifts in a state of two scalars rotating fast around each other, can be compared to the anomalous dimensions of large spin composite operators, due to current and stress tensor respectively, on the CFT side. While the calculations of \cite{Fitzpatrick:2011hh} and \cite{Fitzpatrick:2014vua} entail this feature in some detail, we have managed to extend their work to an $O(2)$ scalar, allowing both gauge and gravitational interactions in composite scalar states. We considered large spin and large twist singlet, traceless symmetric and anti-symmetric composite states, and the results matched with the corresponding anomalous dimensions, computed in the field theory.

\section{Discussion}

\begin{itemize}
\item{ We have analyzed the anomalous dimension of the trace, symmetric-traceless and antisymmetric-traceless large spin operators for the $O(N)$ models.}

\item{The anomalous dimensions have leading twist behaviour in the limit $\ell\gg n\gg1$ which is consistent with the leading twist behaviour given in \cite{Kaviraj:2015cxa} and \cite{Kaviraj:2015xsa}}

\item{In the $O(N)$ model we notice that the effect of the additional minimal twist operators show up in every kind of large spin operators. Thus it is difficult to interpret the monotonicity property of the anomalous dimension. However, demanding monotinicity of the anomalous dimensions might lead to interesting constraints between the OPE squared coefficients for various contributions}

\item{ We have also set up an example holographic verification by considering the $O(2)$ model as a probe on both sides of the duality. We have a large $N$ CFT that allows a holographic dual; the $O(2)$ model is realised through a charged scalar in the bulk, and the gravitational and gauge interactions were used to compute the anomalous dimensions holographically.}

\item{It will be interesting to see the same effects from holographic side by considering the entire $O(N)$ in the probe limit. While it will also be interesting to consider the $O(N)$ model as a standalone theory in the boundary, repeating the bulk calculation for the energy shifts in the bulk will be complicated since now the bulk will be polluted by the predominant higher spin interactions.}

\item{If one considers correlators of spinning fields, one gets different double twist operators \cite{Li2}. Demanding negativity of the anomalous dimensions reproduces the positivity of energy flux in AdS. It will be interesting to study what happens for the higher twist operators of that kind.}
\end{itemize}

\section{Acknowledgements}
We thank Aninda Sinha for discussions and support during the course of this work and also for useful comments during the preparation of the manuscript. We also thank Zohar Komargodski for comments on the draft.

\end{document}